\documentstyle[12pt,psfig,axodraw,a4]{article}
\def\bild#1#2{    
        \vspace*{-5mm}
        \begin{center}
        \begin{math}
        \epsfxsize#2cm
        \epsffile{#1}
        \end{math}
        \end{center}
        }
\newcommand{\vs}{\vspace{-0.25cm}}

\begin{document} 
\begin{center}
\large{\bf Chiral corrections to kaon-nucleon scattering lengths}  

\bigskip 

\bigskip

N. Kaiser\\

\bigskip

Physik Department T39, Technische Universit\"{a}t M\"{u}nchen,\\
    D-85747 Garching, Germany

\end{center}

\bigskip

\begin{abstract}
We calculate the threshold T-matrices of kaon-nucleon and antikaon-nucleon 
scattering to one loop order in SU(3) heavy baryon chiral perturbation theory. 
To that order the complex-valued isospin-1 $\overline KN$ threshold T-matrix 
can be successfully predicted from the isospin-0 and 1 $KN$ threshold 
T-matrices. As expected perturbation theory fails to explain the isospin-0 
$\overline KN$ threshold T-matrix which is completely dominated by the nearby 
subthreshold $\Lambda^*(1405)$-resonance. Cancelations of large terms of 
second and third chiral order are observed as they seem to be typical for
SU(3) baryon chiral perturbation theory calculations. We also give the kaon 
and eta loop corrections to the $\pi N$ scattering lengths and we investigate
$\pi\Lambda$ scattering to one-loop order. The second order s-wave low-energy
constants are all of natural size and do not exceed 1\,GeV$^{-1}$ in magnitude.
   
\end{abstract}

\bigskip
PACS: 12.38.Bx, 12.39.Fe, 13.75.Gx, 13.75.Jz

\bigskip

{\it To be published in: Physical Review C.}

\bigskip 

Chiral symmetry severely constrains the interaction of the strongly interacting
particles at low energies, first formulated in terms of current algebra. One of
the most splendid successes of current algebra was the Tomozawa-Weinberg
prediction \cite{weinb} for the s-wave pion-nucleon scattering lengths
$a^\pm_{\pi N}$. The tool to systematically investigate the consequences of 
spontaneous and explicit chiral symmetry breaking in QCD is chiral perturbation
theory. Observables are calculated with the help of an effective field theory
formulated in terms of the Goldstone bosons ($\pi, K, \overline K, \eta$) and
the  low-lying baryons ($N, \Lambda, \Sigma, \Xi$). A systematic expansion in
small external momenta and meson masses is possible. Pion-loop corrections to
$a^\pm_{\pi N}$  have been first calculated in ref.\cite{bkm}. In the case of 
the isospin-odd $\pi N$-scattering length $a_{\pi N}^-$ the chiral loop 
correction in fact closes the about $15\%$ gap between the lowest order
Tomozawa-Weinberg prediction and the empirical value \cite{psi}. Recently, the
complete fourth order one-loop analysis of low energy $\pi N$-scattering has
been performed in ref.\cite{nadja}.

Kaon-nucleon and antikaon-nucleon scattering are then of interest as a testing
ground of the three-flavor chiral dynamics. In comparison to the SU(2) sector
of $\pi N$-interaction these processes involve several new complications. 
First, the pertinent expansion parameter is much larger since the pion mass
gets replaced by the kaon mass. Secondly, for $\overline KN$-scattering there
exist inelastic channels involving a pion and a hyperon down to threshold,
while $KN$-scattering is purely elastic at low energies. The dynamical
differences in $KN$ versus $\overline K N$-scattering naturally show up in the
values of the s-wave scattering lengths or the equivalent threshold T-matrices,
$T_{KN}^{(0,1)}=4\pi (1+m_K/M_N) a^{(0,1)}_{KN}$ and  $T_{\overline KN}^{(0,1)}
=4\pi(1+m_K/M_N) a^{(0,1)}_{\overline KN}$. The superscripts denote here the
total isospin, $I=0$ or $I=1$. In a combined  dispersion relation analysis of
$KN$ and $\overline K N$-scattering data the following values for the threshold
T-matrices have been obtained by Martin \cite{martin}, 
\begin{equation} T^{(0)}_{K N} =0.4\,{\rm fm}\,, \qquad T^{(1)}_{KN}= -6.3\,
{\rm fm}\,, \end{equation}
\begin{equation} T^{(0)}_{\overline K N} =(-32.6 +13.0\,i)\,{\rm fm}\,, \qquad 
T^{(1)}_{\overline KN}= (7.1+11.5\,i)\, {\rm fm}\,.   \end{equation}
A compilation of values from other determinations can be found in \cite{dumbr}.
We consider the results from the dispersion relation analysis most reliable and
therefore compare the chiral calculation only with the empirical values given 
in eqs.(1,2).

Let us begin with recalling the effective SU(3) chiral Lagrangians for 
meson-meson and meson-baryon interaction, which read at leading order,
\begin{equation} {\cal L}_{\phi\phi}^{(2)} = f^2{\rm tr} \Big( u_\mu u^\mu 
+{\chi_+ \over 4} \Big) \,,  \end{equation}
\begin{equation}
{\cal L}^{(1)}_{\phi B} = {\rm tr}(\overline B(i\partial_0B+[\Gamma_0,B]))-D
\, {\rm tr} (\overline B\{\vec \sigma \cdot \vec u,B\})- F\, {\rm tr} 
(\overline B[\vec \sigma \cdot \vec u , B])\,.\end{equation}
Here, the su(3)-matrix $B$ represents the octet baryon fields $(N,\Lambda,
\Sigma,\Xi)$. The chiral connection $\Gamma^\mu = {i\over 2} [\xi^\dagger, 
\partial^\mu \xi] $ and the axial vector quantity $u^\mu = {i\over 2}
\{\xi^\dagger, \partial^\mu \xi\}$ generate interaction terms with the 
Goldstone bosons $(\pi, K, \overline K, \eta)$. These Goldstone boson fields 
are collected in a SU(3)-matrix $\xi = \exp(i \phi/2f)$. The parameter $f$ 
is the weak meson decay constant in the SU(3) chiral limit. Furthermore, $D$ 
and $F$ in eq.(4) denote the SU(3) axial vector coupling constants of the 
baryons and $\vec \sigma$ is the usual Pauli spin-vector. 
Explicit chiral symmetry breaking is  introduced via the quantity $\chi_+ = 
\xi^\dagger \chi \xi+ \xi \chi \xi^\dagger$ whose vacuum expectation value
$2\chi=2\,{\rm diag}(m_\pi^2,m_\pi^2,2m_K^2-m_\pi^2)$ can be expressed in terms
of the pion and kaon masses squared. At next-to-leading order the part of the 
effective chiral Lagrangian relevant for s-wave meson-baryon interaction reads,
\begin{eqnarray} {\cal L}_{\phi B}^{(2)} &=& b_D\,{\rm tr}(\overline B\{\chi_+,
B\} ) + b_F\, {\rm tr} (\overline B[\chi_+,B]) + b_0\, {\rm tr}\,(\overline B
B)\,{\rm tr} (\chi_+)  \nonumber \\ &&+ \Big( 2d_D+{D^2 -3F^2 \over 2M_0}\Big)
\, {\rm tr} (\overline B \{ u_0^2, B\} ) + \Big(2d_F-{DF\over M_0}\Big) 
\, {\rm tr} (\overline B [u_0^2,B]) \nonumber\\ &&+ \Big( 2d_0 +{F^2-D^2
\over 2M_0} \Big) \, {\rm tr}(\overline BB)\,{\rm tr} (u_0^2) + \Big( 
2d_1+{3F^2 -D^2 \over 3M_0} \Big) \, {\rm tr} (\overline B u_0)\, 
{\rm tr}(u_0 B)\,.  \end{eqnarray}
The first three terms proportional to the low-energy constants $b_{D,F,0}$
are chiral symmetry breaking ones and thus contribute to the mass splittings in
the baryon octet. The remaining double-derivative terms are accompanied by
low-energy constants $d_{D,F,0,1}$ plus $1/M_0$-corrections, where $M_0$ is the
baryon mass in the chiral limit. In the heavy baryon formalism used here such
second order terms with fixed coefficients stem from the $1/M_0$-expansion
\cite{bugra} of the original relativistic leading order Lagrangian ${\cal
L}^{(1)}_{\phi B}$.

Now, we turn to the chiral expansion of the $KN$ and $\overline K N$
scattering threshold T-matrices. At leading order ${\cal O}(q)$ one simply has 
the analog of the Tomozawa-Weinberg relation for kaons and antikaons, 
\begin{equation} T_{KN}^{(0)}=0\,, \qquad T_{KN}^{(1)}=-{m_K \over f_K^2}\,,
\qquad  T_{\overline KN}^{(0)}={3m_K\over 2f_K^2}\,, \qquad T_{\overline KN}^
{(1)}={m_K  \over 2 f_K^2}\,. \end{equation}
We have written these contributions with the physical kaon decay constant 
$f_K$ instead with $f$ (the chiral limit value). Therefore eq.(6) 
subsumes already all those one-loop corrections at ${\cal O}(q^3)$ which 
renormalize $f$ to $f_K$. At next-to-leading order ${\cal O}(q^2)$ one has the
tree-level contributions from the effective Lagrangian ${\cal L}^{(2)}_{\phi 
B}$. These contributions can be compactly written in the form, 
\begin{equation} T_{KN}^{(0)}={m_K^2 \over f_K^2}\, C_0\,, \qquad T_{KN}^{(1)}
={m_K^2 \over f_K^2}\,C_1 \,,\qquad  T_{\overline KN}^{(0)}={m^2_K\over 2f_K^2}
\,(3C_1 -C_0)\,,\qquad T_{\overline KN}^{(1)}={m^2_K \over2f_K^2}\,(C_0+C_1) 
\,,\end{equation}
when introducing the two combinations of low-energy constants $C_{0,1}$ 
relevant for $KN$ and $\overline K N$-scattering,  
\begin{equation}C_0=2d_0-2d_F-d_1+4b_F-4b_0+{D\over 3M_0}(3F-D)\,,
\end{equation}
\begin{equation}C_1=2d_D+2d_0+d_1-4b_0-4b_D-{D^2+3F^2\over6M_0}\,.
\end{equation}
The reason for the appearance of only two independent combinations $C_{0,1}$ 
is crossing symmetry which relates $KN$ and $\overline KN$-scattering (see
eq.(14) below for the general crossing relation). 
  
\bigskip
\bigskip

\bild{knfig1.epsi}{12}

\bigskip

\bild{knfig2.epsi}{12}

\bigskip
\hspace{5.6cm} {\bf *}
\bild{knfig3.epsi}{12}

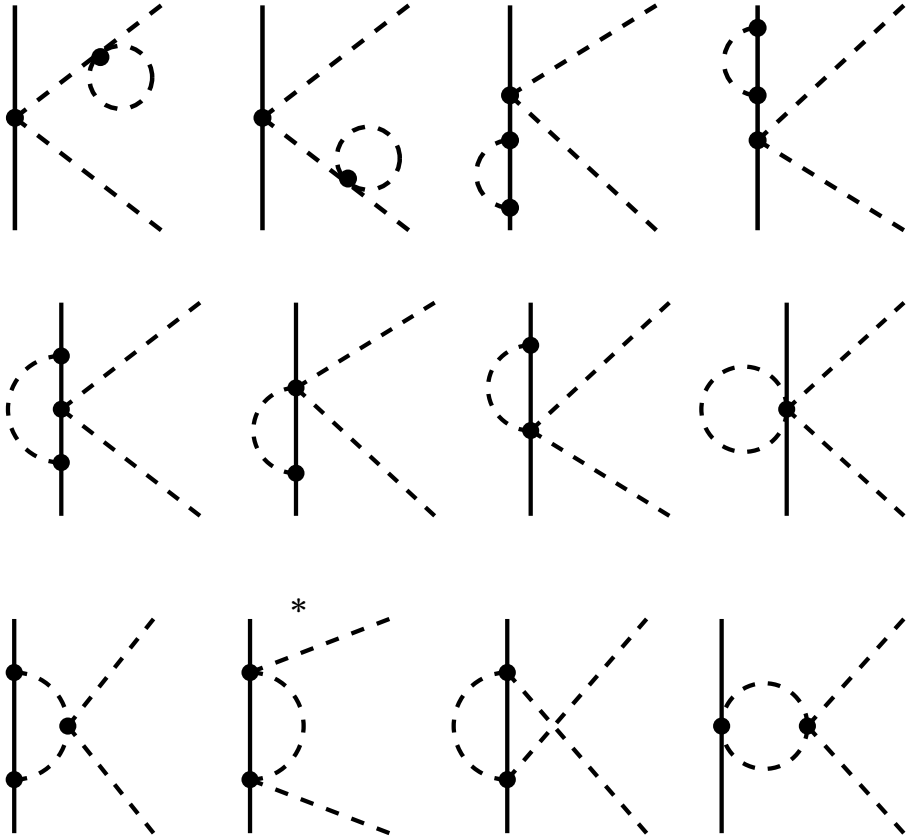
{\it Fig.\,1: Nonvanishing one-loop diagrams for meson-baryon scattering at
threshold. Dashed lines represent Goldstone bosons ($\pi, K,\overline K,\eta$) 
and full lines represent baryons ($N,\Lambda, \Sigma,\Xi$). The nonvanishing
imaginary parts of $T_{\overline KN}^{(0,1)}$ come exclusively from the diagram
marked by an asterisk.}

\bigskip 

At next-to-next-to-leading order ${\cal O}(q^3)$ one has the contribution from
all one-loop graphs generated by the vertices of ${\cal L}^{(2)}_{\phi\phi
}$ and ${\cal L}^{(1)}_{\phi B}$. In case of the threshold T-matrices the
loop calculation simplifies considerably since all those diagrams in which the
in-  or out-going meson couples directly to the baryon line vanish identically 
($\vec\sigma \cdot \vec q =0$). The remaining set of non-vanishing one-loop
diagrams is shown in Fig.\,1. We use dimensional regularization and minimal
subtraction to evaluate divergent loop integrals (for details see appendix B of
ref.\cite{review}). After renormalizing $f$ to $f_K$ in the leading order terms
(as done already in eq.(6)) and summing up the contributions from all diagrams
in Fig.\,1 one finds the following renormalized one-loop chiral corrections to
the $KN$ and $\overline K N$ threshold T-matrices,    
\begin{eqnarray} T_{KN}^{(0)}&=&{m_K^2\over 16\pi^2 f_\pi^2 f_K^2}\bigg\{3 m_K
\ln{m_\pi \over |m_K|}+ 3\sqrt {m_K^2-m_\pi^2} \,\ln{m_K+\sqrt{m_K^2-
m_\pi^2} \over m_\pi}  \nonumber\\ && + 
\pi (D-3F) \bigg[(D+F) {m_\pi^2 \over m_\eta+m_\pi} +(7D+3F) {m_\eta \over 6}
\bigg] \bigg\} \,, \end{eqnarray} 
\begin{eqnarray} T_{KN}^{(1)}&=&{m_K^2\over 16\pi^2 f_\pi^2 f_K^2}\bigg\{m_K 
\bigg(2\ln{m_\pi \over\lambda} + \ln {|m_K| \over \lambda}+3 \ln{m_\eta \over 
\lambda} -3\bigg)  \nonumber \\ && +2\sqrt{m_K^2- m_\pi^2}\, \ln{m_K+\sqrt
{m_K^2-m_\pi^2} \over m_\pi} -3\sqrt{m_\eta^2-m_K^2}\,\arccos{m_K \over m_\eta}
\nonumber  \\  && + {\pi\over 6} (3F-D)\bigg[ 2(D+F) {m_\pi^2 \over
m_\eta+m_\pi} +(D+5F) m_\eta  \bigg] \bigg\} \,,
\end{eqnarray} 
\begin{eqnarray} T_{\overline KN}^{(0)}&=&{m_K^2 \over 32\pi^2 f_\pi^2 f_K^2}
\bigg\{ 3m_K \bigg(3- \ln{m_\pi \over \lambda}-2\ln {|m_K| \over \lambda} -3 
\ln{m_\eta \over \lambda} \bigg) \nonumber\\ &&+3\sqrt {m_K^2-m_\pi^2} \bigg(
i\,\pi -\ln{m_K+\sqrt{m_K^2-m_\pi^2} \over m_\pi} \bigg) -9
\sqrt{m_\eta^2-m_K^2} \arccos{-m_K\over m_\eta} \nonumber \\ && + 
\pi (3F-D) \bigg[2(D+F) {m_\pi^2 \over m_\eta+m_\pi} +(5D+9F) {m_\eta \over 3}
\bigg] \bigg\} \,, \end{eqnarray} 
\begin{eqnarray} T_{\overline KN}^{(1)}&=&{m_K^2\over 32\pi^2 f_\pi^2 f_K^2}
\bigg\{ m_K \bigg(3-5\ln{m_\pi \over\lambda} +2 \ln {|m_K| \over \lambda}-3 
\ln{m_\eta \over \lambda} \bigg)\nonumber  \\  && +5\sqrt{m_K^2-m_\pi^2}\bigg(
i \,\pi - \ln{m_K+\sqrt{m_K^2-m_\pi^2} \over m_\pi} \bigg)-3\sqrt{m_\eta^2-
m_K^2}\, \arccos{ -m_K \over m_\eta}  \nonumber \\ && + {\pi\over 3} (D-3F)
\bigg[ 2(D+F) {m_\pi^2 \over m_\eta+m_\pi} +(3D-F)m_\eta  \bigg] \bigg\} \,,
\end{eqnarray} 
with $\lambda\sim 1$\,GeV the scale parameter introduced in dimensional 
regularization. Note that we have used the symmetrical product $f_\pi^2 
f_K^2$ in the denominator of the prefactor in order to be more realistic on the
rescattering process $\overline K N\to (\pi \Lambda, \pi\Sigma) \to \overline K
N$ which generates imaginary parts. The corresponding diagram is the one 
marked by an asterisk  in Fig.\,1. The total sums of all (renormalized)
one-loop corrections eqs.(10,11,12,13) have remarkable properties which do not
hold for the contributions from individual graphs. First, $m_K^2$ factors
out. Secondly, the prefactors of chiral logarithms $\ln(m_{\pi,K,\eta}/
\lambda)$ are independent of the axial vector coupling constants $D$ and
$F$. Thirdly, the terms bilinear in $D$ and $F$ are finite, they carry an
extra factor of $\pi$, and moreover $D-3F$, which is proportional to the $\eta 
NN$-coupling constant, can be factored out of these contributions. The finite 
terms bilinear in $D$ and $F$ have been evaluated earlier in ref.\cite{lee}. 
Because of these properties there are only two chiral families of graphs in 
Fig.\,1. Family\,1 is independent of $D$ and $F$ and it consists of the 
diagrams generated (exclusively) by meson-baryon vertices with an even number 
of meson-lines. Family\,2 on the other hand is bilinear in $D$ and $F$ and it 
comprises the diagrams generated (in addition) by meson-baryon vertices with 
an odd  number of meson-lines. Note  also that the loop contributions
eqs.(10,11,12,13) remain nonsingular in the SU(2) chiral limit $m_\pi\to 0$ and
$m_{K,\eta}$ fixed. In addition to the contributions from one-loop graphs one
has also further counterterm contributions at order ${\cal O}(q^3)$ which
balance the scale dependence of the chiral logarithms
$\ln(m_{\pi,K,\eta}/\lambda)$. In the 
case of threshold $\pi N$-scattering \cite{bkm}  the third order counterterm
contribution has been estimated from resonance exchange and it was found to be
much smaller than the chiral loop contribution. We assume that similar features
hold for $KN$ and $\overline K N$-scattering.  Note also that the low-energy
constants $d_{D,F,0,1}$ of the second order SU(3) chiral Lagrangian are not
well determined at present. Let us finally mention the crossing symmetry
relations which allow to calculate the $\overline KN$ threshold T-matrices
directly from the $KN$ threshold T-matrices by changing the sign of the kaon
threshold energy $m_K$,   
\begin{equation} T_{\overline KN}^{(0)} = {1\over 2} \Big[ 3T_{KN}^{(1)}-
T_{KN}^{(0)}\Big]_{m_K\to -m_K} \,, \qquad T_{\overline KN}^{(1)} = {1\over 2}
\Big[ T_{KN}^{(0)}+ T_{KN}^{(1)}\Big]_{m_K\to -m_K} \,. \end{equation}
Obviously, these relations apply order by order in the chiral expansion. 

As a further application of the full SU(3) scheme we reconsider the chiral
expansion of the isospin-even and isospin-odd $\pi N$-scattering threshold 
T-matrices $T^{\pm}_{\pi N}=4\pi  (1+m_\pi/M_N) a^{\pm}_{\pi N}$. The
contributions at order ${\cal O}(q)$ and ${\cal O}(q^2)$ appear selectively in
the isospin-odd and isospin-even amplitude as,  
\begin{equation} T_{\pi N}^-={m_\pi \over 2f_\pi^2}\,, \qquad T_{\pi N}^+ = 
{m_\pi^2 \over f_\pi^2}\bigg\{ d_D+d_F +2 d_0 
-2b_D-2b_F-4b_0-{(D+F)^2 \over 4M_0}\bigg\} \,. \end{equation} 
Here, we have again subsumed in $T_{\pi N}^-$ already those loop corrections 
which renormalize $f$ to $f_\pi$, the physical pion decay constant. The
remaining $(\pi,K,\eta)$-loop contributions at order ${\cal O}(q^3)$ generated
by the diagrams in Fig.\,1 are of the form, 
\begin{equation} T^+_{\pi N} = {3m_\pi^2 \over 64 \pi f_\pi^4} \bigg\{ (D+F)^2
m_\pi -2 \sqrt{m_K^2-m_\pi^2 } -(D-3F)^2 {m_\eta \over 9} \bigg\} \,,
\end{equation}
\begin{equation} T^-_{\pi N} = {m_\pi^2 \over 16\pi^2 f_\pi^4} \bigg\{ m_\pi
\bigg( {3\over 2} -2\ln{m_\pi \over \lambda}-\ln {m_K \over \lambda}\bigg)
-\sqrt{m_K^2-m_\pi^2} \,\arcsin{m_\pi \over m_K} \bigg\} \,. \end{equation}   
Note that the kaon and eta-loop contributions to the isospin-even threshold
T-matrix $T^+_{\pi N}$ start out at order $m_\pi^2$ and therefore would have to
be accounted for by the second order low-energy constants $c_{1,2,3}$ 
\cite{bkm} in a reduction to SU(2). The isospin-odd threshold T-matrix 
$T^-_{\pi N}$ on the other hand receives an additional small kaon-loop 
contribution proportional to $-1/2-\ln( m_K/\lambda)$  which will be viewed in
a SU(2) reduction as a part of the third order low-energy constant 
$B^r(\lambda)$ (see ref.\cite{bkm}).  

Finally, we consider low-energy elastic $\pi \Lambda$-scattering in SU(3) 
baryon chiral perturbation theory. With only one total isospin-state and 
inherent crossing symmetry the $\pi\Lambda$-system is particularly simple. For
the $\pi\Lambda$ threshold T-matrix $T_{\pi\Lambda}$ the leading order ${\cal
O}(q)$ contribution vanishes. The second order ${\cal O}(q^2)$ contribution 
from the effective Lagrangian ${\cal L}_{\phi B}^{(2)}$ and the complete ${\cal
O}(q^3)$ chiral loop corrections (see Fig.\,1) read together,
\begin{equation} T_{\pi\Lambda} = {2m_\pi^2 \over 3f_\pi^2}\bigg\{d_D+3d_0-2b_D
-6b_0-{D^2 \over 2M_0}\bigg\} +{m_\pi^2 \over 16 \pi f_\pi^4} \bigg\{ D^2 \Big(
 m_\pi-{m_\eta \over 3} \Big) -3\sqrt{m_K^2-m_\pi^2}\bigg\} \,.\end{equation}
We will use this expression to estimate the $\pi\Lambda$ phase-shift at the
$\Xi$-mass, a quantity which has recently received some interest 
\cite{pilam1,pilam2}. 
  
Next, we turn to the discussion of numerical results. We use for the meson
masses $m_\pi = 139.57\,$MeV, $m_K=493.68\,$MeV and $m_\eta=\sqrt{(4m_K^2-
m_\pi^2)/3}= 564.33\,$MeV (the value implied by the GMO-relation which 
deviates only by $3.1\%$ from the physical $\eta$-mass $m_\eta = 547.3\,$MeV
\cite{pdg}). The pion and kaon decay constants $f_\pi= 92.4\, $MeV and $f_K =
113\,$MeV \cite{pdg} are well-known and for the baryon axial vector coupling 
constants we use $D=0.8$ and $F=0.5$. This choice corresponds via the
Goldberger-Treiman relation to a strong $\pi NN$-coupling constant of $g_{\pi
N}=(D+F)M_N/f_\pi = 13.2$ which is consistent with present empirical
determinations \cite{pavan}. The ratio $D/F=1.6$ agrees with the one extracted
from semileptonic hyperon decays \cite{bourquin}.  

In order to make predictions and to determine
remaining low-energy constants we do the following. The empirical value of the
isospin-odd $\pi N$ threshold T-matrix $T^-_{\pi N}=(1.847\pm 0.086)$\,fm
\cite{psi} is reproduced at one-loop order (eqs.(15,17)) with the 
renormalization scale $\lambda$ set equal to $\lambda = 0.95\,$GeV. We adopt
then this value of $\lambda$ also for the $KN$ and $\overline K N$  threshold 
T-matrices. The combinations of low-energy constants $C_{0,1}$ are adjusted 
such that the empirical values of $T^{(0,1)}_{KN}$ (eq.(1)) are reproduced by 
their  chiral expansions up to and including order ${\cal O}(q^3)$
(eqs.(6,7,10,11)). Inserting the resulting values $C_0 = 0.12\,$fm\,$=0.60\,
$GeV$^{-1}$ and $C_1 = 0.41\, $fm\,$=2.09\,$GeV$^{-1}$ into the formulas for 
$\overline KN$ threshold T-matrices (eqs.(6,7,12,13)) one predicts this way,
\begin{equation} T^{(0)}_{\overline K N} =(30.4 +6.2\,i)\,{\rm fm}\,, \qquad 
T^{(1)}_{\overline KN}= (7.1+10.4\,i)\, {\rm fm}\,.   \end{equation}
The complex-valued isospin-1 amplitude $T^{(1)}_{\overline KN}$ is in good 
agreement with the corresponding empirical value eq.(2), only its imaginary
comes out about $10\%$ too small.\footnote{Note that relativistic 
$1/M_0$-corrections to the (complex-valued) one-loop amplitudes are considered
as part of the higher order terms at ${\cal O}(q^4)$ in the heavy baryon
formalism used here. A reevaluation of chiral corrections to $KN$ and
$\overline KN $-scattering in a fully relativistic framework such as the one of
ref.\cite{becher} can tell how important these corrections are for both the 
real and the imaginary part. Since the ratio $m_K/M_0\simeq 0.5$ is not that
small, the effects could be sizeable.}  However, one-loop chiral perturbation
theory fails completely in case of the isospin-0 amplitude  $T^{(0)}_{\overline
KN}$. The predicted  real part has the right magnitude but the wrong positive
sign and the imaginary is about a factor 2 too small. This should not come as a
surprise since the isospin-0 $\overline KN$ s-wave channel is known to be 
completely dominated by the nearby subthreshold $\Lambda^*(1405)$-resonance. 
Only when employing non-perturbative methods \cite{siegel,oset,oller} one can 
generate the $\Lambda^*(1405)$-resonance as a quasi-bound $\overline KN$-state 
from the lowest order attractive chiral meson-baryon interaction in this 
channel. It is also not meaningful to add to the isospin-0 $\overline K N$ 
amplitude of one-loop chiral perturbation theory an explicit 
$\Lambda^*(1405)$-resonance contribution as done in \cite{lee}. 

The empirical value of the scattering length $a_{\overline KN}^{(0)}= (-1.70 
+0.68\,i)\,$fm \cite{martin} and the presence of the subthreshold $\Lambda^*
(1405)$-resonance are in fact intimately related with each other. In order to
demonstrate this we show in Fig.\,2 the real and imaginary part of the 
isospin-0 $\overline KN$ s-wave scattering amplitude in "unitarized" scattering
length approximation \cite{queen}, 
\begin{equation} f_{\overline KN}^{(0)} (W) = \Big[\Big(a_{\overline KN}^{(0)}
\Big)^{-1} - i\, Q_* \Big]^{-1}  \,, \quad 2Q_* = \sqrt{
W^2-2(M_N^2+m_K^2)+(M_N^2-m_K^2)^2W^{-2} }\,, \end{equation}
as a function of the $\overline K N$ invariant mass $W$. The prominent
resonance structure following already from the simple parametrization eq.(20) 
is clearly visible in Fig.\,2. The position and slope of the zero-crossing of 
the real part Re\,$f_{\overline KN}^{(0)}(W)$ suggest a resonance mass of $W_0=
1416.4$\,MeV and a width of $\Gamma= 25.2\,$MeV. Such values of the mass and 
width are typically found for the $\Lambda^*(1405)$-resonance in extrapolations
below the  $\overline K N$  threshold \cite{pdg}.

\bigskip
\bigskip

\bild{lam1405.epsi}{12}

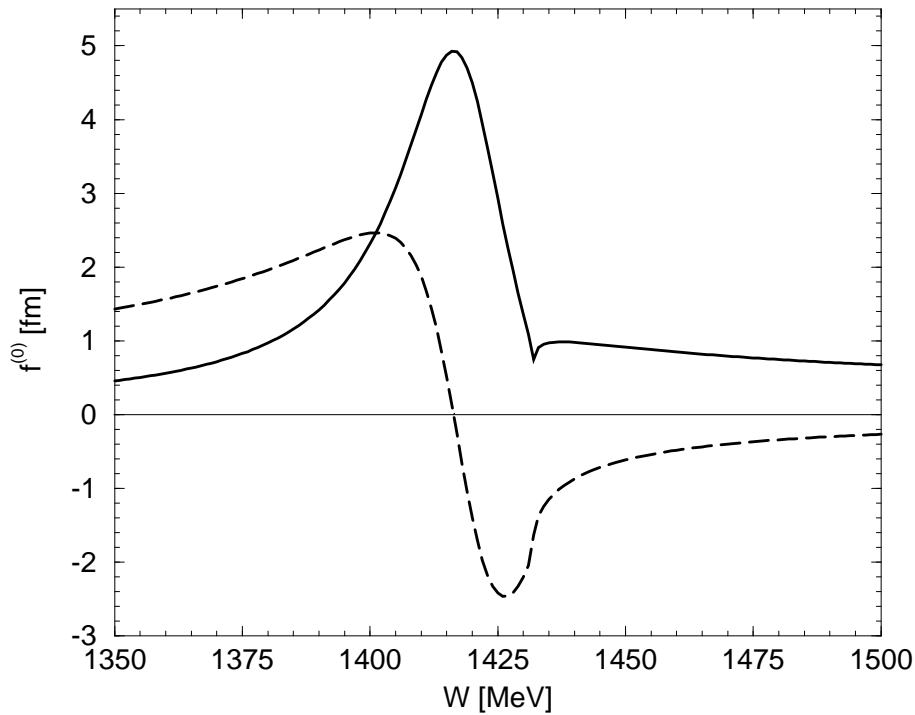
{\it Fig.\,2: The isospin-0 $\overline KN$ s-wave scattering amplitude in
"unitarized" scattering length approximation versus the $\overline KN$ 
invariant mass $W$. The real part (imaginary part) of $f^{(0)}_{\overline
KN}(W)$ is shown by the dashed (solid) line.}  

\bigskip

Let us now take a closer look at the chiral expansion of $T^{(0,1)}_{KN}$ and 
$T^{(0,1)}_{\overline KN}$. Numerically, one finds for the contributions at  
first, second and third chiral order: $T^{(0)}_{KN}= (0+2.29 -1.87)\,$fm, 
$T^{(1)}_{KN} = (-7.63 +7.83 -6.54)\,$fm,  $T^{(0)}_{\overline KN}= (11.44+
10.59+8.35 + 6.23\,i)\,$fm and $T^{(1)}_{\overline KN}= (3.81+5.06-1.74+10.39
\,i)\,$fm. One observes cancelations of large contributions at second and third
chiral order, in particular for $T^{(0,1)}_{KN}$. This feature has also 
occurred in the chiral expansion of the baryon masses and magnetic moments
performed to fourth order in ref.\cite{bugra,sven}. Such cancelations of large
terms seem to be generic for SU(3) baryon chiral perturbation theory 
calculations.

Finally, we want to use the present novel one-loop results to extract values 
of the s-wave low-energy constants $d_{D,F,0,1}$. The three chiral symmetry
breaking parameters $b_{D,F,0}$ and $M_0$ can be determined from the octet
baryon masses and the $\pi N$ sigma-term $\sigma_{\pi N}(0)= 45\,$MeV (the
central value of ref.\cite{gls}). Using the one-loop expressions given in
chapter 6.1 of ref.\cite{review} and inserting $f=f_\pi$ in ($\pi,\eta$)-loops
and $f=f_K$ in kaon-loops one finds in a best fit $b_D= 0.042\,$GeV$^{-1}$,
$b_F= -0.557\,$GeV$^{-1}$, $b_0= -0.789\,$GeV$^{-1}$ and $M_0=918.4\,$MeV. With
the additional piece of information given by the isospin-even $\pi N$ threshold
T-matrix $T^+_{\pi N} = (-0.045\pm 0.088) \,$fm \cite{psi} and the values 
$C_0= 0.6\,$GeV$^{-1}$ and $C_1= 2.1\,$GeV$^{-1}$ deduced from the $KN$ 
threshold T-matrices one obtains $d_F=-0.968\,$GeV$^{-1}$, $2d_0+d_D= -1.562\,
$GeV$^{-1}$ and $d_1+d_D= 1.150\,$GeV$^{-1}$. These numbers are comparable both
in magnitude and sign with the ones of ref.\cite{pwave}. There a large amount
of pion and photon induced $(\eta, K)$-production data has been fitted in a 
coupled channel approach including s- and p-waves. If we furthermore assume 
that all second  order low-energy constants do not exceed 1\,GeV$^{-1}$ in
magnitude (for which there is no known exception so far) we can make
a prediction for the $\pi\Lambda $ s-wave phase-shift at the $\Xi$-mass. Using 
the scattering length approximation  $\delta_{\pi \Lambda} = q_\pi^{cm} T_{\pi 
\Lambda}[4\pi(1+m_\pi/M_\Lambda)]^{-1}$ with $q_\pi^{cm}=139\,$MeV  and the
one-loop expression eq.(18) for $T_{\pi \Lambda}$ we obtain with $d_0 \simeq 
-1.0\,$GeV$^{-1}$ \cite{pwave} a $\pi \Lambda $ threshold T-matrix of $T_{\pi 
\Lambda} \simeq -1.0$\,fm and a phase-shift at the $\Xi$-mass of $\delta_{\pi 
\Lambda}\simeq-2.8^\circ$. This is considerably
smaller than the K-matrix result $\delta_{\pi \Lambda} \simeq -7^\circ$ of
ref.\cite{pilam1}. The present perturbative one-loop result is however outside
the very small range $0^\circ \leq \delta_{\pi \Lambda} \leq 1.1^\circ$ found
in the relativistic chiral coupled channel approach of ref.\cite{pilam2}. 

In summary, we have calculated here in SU(3) heavy baryon chiral perturbation 
theory the threshold T-matrices for $KN$, $\overline K N$, $\pi N$ and $\pi
\Lambda$ scattering to one-loop order. The kaon-loop contribution to the
isospin-odd amplitude $T^-_{\pi N}$ is small. The complex-valued isospin-1
amplitude $T^{(1)}_{\overline KN}$  can be successfully predicted from the two
$KN$ amplitudes $T^{(0,1)}_{KN}$. As expected perturbation theory fails in the
isospin-0 $\overline K N$ channel which is completely dominated by the nearby 
subthreshold $\Lambda^*(1405)$-resonance. Cancelations of large terms at 
second and third order seem to be a generic feature of SU(3) baryon chiral
perturbation theory calculations. All second order low-energy constants are of
natural size and do not exceed 1\,GeV$^{-1}$ in magnitude. At one-loop order we
predict the $\pi \Lambda$ phase-shift at the $\Xi$-mass to be $\delta_{\pi
\Lambda} \simeq -2.8^\circ$. In order to quantify the corrections to the 
present one-loop results fully relativistic calculations and higher order 
${\cal O}(q^4)$ calculations in the heavy baryon framework of SU(3) chiral
perturbation theory are needed.


\begin{thebibliography}{99}
\bibitem{weinb} Y. Tomozawa, {\it Nuovo Cimento} {\bf 46A} (1966) 707; 
S. Weinberg, {\it Phys. Rev. Lett.} {\bf 17} (1966) 616.\vs 
\bibitem{bkm} V. Bernard, N. Kaiser and Ulf-G. Mei{\ss}ner, {\it
Phys. Lett.} {\bf B309} (1993) 421; {\it Phys. Rev.} {\bf C52} (1995) 2185.\vs
\bibitem{psi} H.Ch. Schr\"oder et al., {\it Phys. Lett.} {\bf B469} (1999) 
25.\vs
\bibitem{nadja} N. Fettes and Ulf-G. Mei{\ss}ner, {\it Nucl. Phys.} {\bf A676}
(2000) 311; and refs. therein.\vs
\bibitem{martin} A.D. Martin, {\it Nucl. Phys.} {\bf B179} (1981) 33.\vs
\bibitem{dumbr} O. Dumbrajs et al., {\it Nucl. Phys.} {\bf B216} (1983) 277.\vs
\bibitem{bugra} B. Borasoy and Ulf-G. Mei{\ss}ner, {\it Ann. Phys. (NY)}
{\bf 254} (1997) 192.\vs
\bibitem{review} V. Bernard, N. Kaiser and Ulf-G. Mei{\ss}ner, {\it
Int. J. Mod. Phys.} {\bf E4} (1995) 193.\vs
\bibitem{lee} C.H. Lee et al., {\it Phys. Lett.} {\bf B326} (1994) 14.\vs   
\bibitem{pilam1} J. Tandean, A.W. Thomas and G. Valencia, hep-ph/0011214.\vs
\bibitem{pilam2} Ulf-G. Mei{\ss}ner and J.A. Oller, hep-ph/0011293.\vs 
\bibitem{pdg} Particle Data Group, D.E. Groom et al., {\it Eur. Phys. J.} {\bf
C15} (2000) 1.\vs
\bibitem{pavan} M.M. Pavan et al., {\it Physica Scripta} {\bf T87} (2000)
65.\vs 
\bibitem{bourquin} M. Bourquin et al., {\it Z. Phys.} {\bf C21} (1983) 27.\vs 
\bibitem{becher} T. Becher and H. Leutwyler, {\it Eur. Phys. J.} {\bf C9}
(1999) 643; and refs. therein.\vs 
\bibitem{siegel} N. Kaiser, P.B. Siegel and W. Weise, {\it Nucl. Phys.} {\bf
A594} (1995) 325.\vs
\bibitem{oset} E. Oset and A. Ramos, {\it Nucl. Phys.} {\bf A635} (1998) 99.\vs
\bibitem{oller} J.A. Oller and Ulf-G. Mei{\ss}ner, {\it Phys. Lett.} {\bf B500}
(2001) 263.\vs
\bibitem{queen} N.M. Queen and G. Violini, "Dispersion theory in high energy
physics" MacMillan Press, London, 1974, page 92.\vs
\bibitem{sven} S. Steininger and Ulf-G. Mei{\ss}ner, {\it Nucl. Phys.}
{\bf B499} (1997) 359.\vs
\bibitem{gls} J. Gasser, H. Leutwyler and M.E. Sainio, {\it Phys. Lett.} {\bf
B253} (1991) 252.\vs 
\bibitem{pwave} J. Caro Ramon, N. Kaiser, S. Wetzel and W. Weise, {\it
Nucl. Phys.} {\bf A672} (2000) 249.\vs
\end{thebibliography}
\end{document}